\documentclass[a4paper,11pt,twoside]{article}

\usepackage{hyperref}
\usepackage[latin1]{inputenc}
\usepackage[T1]{fontenc}
\usepackage{a4wide}
\usepackage{amsmath,amsthm}
\usepackage{amssymb}
\usepackage{a4wide}
\usepackage{graphicx,epsfig}
\usepackage{color}
\usepackage{epic}
\usepackage{enumerate}

\newcommand{\dps}{\displaystyle }
\newcommand{\beqn}{\begin{eqnarray*}}
\newcommand{\eeqn}{\end{eqnarray*}}

\newcommand{\ri}{{\rm i}}
\newcommand{\lH}{\widetilde{H}}
\newcommand{\lA}{\widetilde{A}}
\newcommand{\lP}{\widetilde{P}}
\newcommand{\lE}{\widetilde{E}}
\newcommand{\lK}{\widetilde{K}}
\newcommand{\OO}{{\rm O}}
\newcommand{\UA}{U_{\rm A}}
\newcommand{\UAe}{U_{\eps, {\rm A}}}
\newcommand{\wP}{\widetilde{P}}
\newcommand{\UAI}{U_{{\rm A},{\rm int}}}

\newcommand{\HA}{H_{\rm A}}

\newcommand{\id}{{\mathbb{I}}}
\newcommand{\rme}{{\rm e}}
\newcommand{\eps}{\varepsilon}

\newcommand{\cE}{{\cal E}}
\newcommand{\cP}{{\cal P}}

\newcommand{\bra}{\left \langle \, }
\newcommand{\ket}{\, \right \rangle}
\newcommand{\lsep}{\, \left | \,}
\newcommand{\rsep}{\, \right | \,}

\newcommand{\R}{\mathbb{R}}

\def\({\left(}
\def\){\right)}
\def\<{\left\langle}
\def\>{\right\rangle}

\newcommand{\1}{\mathbb{I}}

\newcommand{\ie}{{\sl i.\,e. }}

\DeclareMathOperator{\Ran}{Ran}

\numberwithin{equation}{section}
\newtheorem{theoreme}{Theorem}
\newtheorem{proposition}[theoreme]{Proposition}

\newtheorem{lemma}[theoreme]{Lemma}

\newtheorem{definition}[theoreme]{Definition}
\newtheorem{assumption}[theoreme]{Assumption}
\newtheorem{remark}[theoreme]{Remark}

\begin{document}

%
%

\title{Gell-Mann and Low formula\\ for degenerate unperturbed states}
\author{
Christian Brouder$^1$, Gianluca Panati$^2$, and Gabriel Stoltz$^3$ \\
  \normalsize
  $^1$: Institut de Min\'eralogie et de Physique des Milieux Condens\'es, \\
  \normalsize
  CNRS UMR 7590, Universit\'es Paris 6 et 7, IPGP, 140 rue de Lourmel,
  75015 Paris, France. \\
  \normalsize
  $^2$: Dipartimento di Matematica, Universit\`a di Roma La Sapienza, Roma, Italy \\
  \normalsize
  $^3$: Universit\'e Paris Est, CERMICS, Projet MICMAC ENPC - INRIA, \\
  \normalsize
  6 \& 8 Av. Pascal, 77455 Marne-la-Vall\'ee Cedex 2, France
}

\maketitle

\begin{abstract}
The Gell-Mann and Low switching allows to transform eigenstates of
an unperturbed Hamiltonian $H_0$ into eigenstates of the modified
Hamiltonian $H_0 + V$. This switching can be performed when the
initial eigenstate is not degenerate, under some gap conditions with
the remainder of the spectrum. We show here how to extend this
approach to the case when the ground state of the unperturbed
Hamiltonian is degenerate. More precisely, we prove that the
switching procedure can still be performed when the initial states
are eigenstates of the finite rank self-adjoint operator $\cP_0 V
\cP_0$, where $\cP_0$ is the projection onto a degenerate
eigenspace of $H_0$.
\end{abstract}

%
%

\section{Introduction}

Adiabatic switching is a crucial ingredient of many-body theory. It
provides a way to express the eigenstates of a Hamiltonian $H_0+V$
in terms of the eigenstates of $H_0$. Its basic idea is to switch
very slowly the interaction $V$, \ie to transform $H_0+V$ into a
time-dependent Hamiltonian of the typical form $H_0 +
\rme^{-\varepsilon|t|} V$, where the small parameter $\varepsilon >
0$ eventually vanishes. It may be expected that an eigenstate of
$H_0+V$ is obtained by taking the limit of an eigenstate of $H_0$,
evolved according to the time-dependent Hamiltonian $H_0 +
\rme^{-\varepsilon|t|} V$ when $\varepsilon$ tends to zero. It turns
out that this naive expectation is not justified since the time-dependent
eigenstate has no limit when $\varepsilon \to 0$ because of some
non-convergent phase factor. When the initial state belongs to a non
degenerate eigenspace, Gell-Mann and Low solved the problem by
dividing out the oscillations by a suitable factor~\cite{GML51}. The
ratio becomes, in the limit $\eps \to 0$, the Gell-Mann and Low
wavefunction. Mathematically, the convergence of this procedure has
been proved in 1989 by Nenciu and Rasche \cite{NR89}, elaborating on
the adiabatic theorem \cite{BF28, Kato50, Garrido64}.

\medskip

On the other side, the physics community realized about fifty years
ago \cite{BH58} that a generalization of the Gell-Mann and Low
formula is needed in the case of a degenerate eigenvalue of $H_0$.
This happens in many practical situations, for instance when the
system contains unfilled shells. This problem has been discussed in
several fields, including nuclear physics, solid state physics,
quantum chemistry and atomic physics, see the references
in~\cite{BSP08,BPS09}. In most cases, it is assumed that there is
\emph{some} eigenstate in the degenerate eigenspace $\cE_0$ of $H_0$
for which the Gell-Mann and Low formula holds. In general however,
the Gell-Mann and Low formula is not applicable when this state is
chosen at random in the degenerate subspace, as illustrated in the
simple model analytically studied in~\cite{BSP08}.

We show in this paper that the switching can be performed provided
the initial eigenstates are also  eigenstates of $\cP_0 V \cP_0 \big
|_{\cE_0}$, the perturbation restricted to act on the degenerate
eigenspace. If the latter operator has itself degenerate eigenvalues,
a further analysis is required, as discussed in Section
\ref{sec:extension}. The result is based on the recent progress in
the mathematical analysis of adiabatic problems, see
\cite{Nenciu80,ASY87,Hagedorn89,Sjostrand1993,HagedornJoye2001,
MartinezSordoni2002,PanatiSpohnTeufel2003,Teufel,HagedornJoye_review} and references therein.

The physical consequences of our result are discussed in the
companion physics paper~\cite{BPS09}, where we also comment on the
formal relation with different types of Green functions.

\medskip

\noindent \textbf{Acknowledgements:} G.P. is grateful to S. Teufel
and J. Wachsmuth for a useful discussion in a preliminary stage of
this work. We gratefully thank an anonymous referee for useful
comments and remarks, which encouraged us to generalize the result
appearing in the first version of the paper.

%
%

\section{Statement of the results}
\label{sec:statement}

\subsection{Spectral structure of the problem}

Consider a Hilbert space $\mathcal{H}$,
a self-adjoint operator $H_0$,
with dense domain $D(H_0) \subset \mathcal{H}$,
and a symmetric perturbation $V$, $H_0$-bounded with relative bound $a < 1$.
Then, according to the Kato-Rellich theorem (Theorem X.12 in~\cite{ReedSimon2}),
$H_0 + \lambda V$ is self-adjoint
on $D(H_0)$ for any $0 \leq \lambda \leq 1$.
We denote\footnote{
  For reasons that will become clear once a time variable is
  introduced, we will always denote with a \, $\widetilde{}$ \ functions
  of the variable $\lambda \in [0,1]$. Untilded functions will have time
  as an argument.
}
\[
\lH(\lambda) = H_0 + \lambda V,
\]
with $\lambda \in [0,1]$.
In all this study, we assume that the spectrum has the following structure.

\begin{assumption}[Structure of the spectrum]
  \label{ass:structure}
  The spectrum of $\lH(\lambda) = H_0 + \lambda V$,
  $\lambda \in [0,1]$, consists of two disconnected pieces
  \[
  \sigma\big(\lH(\lambda)\big) = \sigma_{N}(\lambda) \cup \Big (
  \sigma\big(\lH(\lambda)\big) \backslash \sigma_{N}(\lambda) \Big )
  \]
  where $\sigma_{N}(\lambda)$ is a finite subset of the discrete spectrum:
  \[
  \sigma_{N}(\lambda) = \Big \{ \lE_j(\lambda), \ j = 1,\dots,N \Big \}
  \subset \sigma_{\rm disc}\left(\lH(\lambda)\right),
  \]
  and the initial state is degenerate: $\lE_j(0) = \lE_k(0)$ for
  all $1 \leq j,k \leq N$.
\end{assumption}

In order to apply results and techniques from adiabatic
theory~\cite{BF28,Kato50,Nenciu80,ASY87},
we make the following standard assumption on the existence of a gap in the spectrum.

\begin{assumption}[Gap condition]
  \label{ass:global_gap}
  There is a gap between the two parts of the spectrum, in the sense
  that:
  \[
  \Delta (\lambda) = \min_{j = 1,\dots,N} \left(
  \min \Big \{ \left|\lE_j(\lambda)-E\right|, \
  E \in \sigma(H(\lambda)) \backslash \big\{ \lE_1(\lambda),\dots,\lE_N(\lambda)\big\} \Big \}
  \right),
  \]
  is bounded from below by a positive constant:
  \[
  \inf_{\lambda \in [0,1]} \Delta(\lambda) = \Delta^\ast > 0.
  \]
\end{assumption}

The projectors associated with the $N$ eigenvalues $\lE_j(\lambda)$
(counted with their multiplicities) are denoted by $\lP_j(\lambda)$,
for $1 \leq j \leq M$ with $M \leq N$.
The projector onto the subspace orthogonal to the eigenspace spanned
by the $N$ eigenvectors is $\lP_{N+1}(\lambda) = \1 - \sum_{j=1}^M
\lP_j(\lambda)$. We denote in the sequel
\[
\cP_0 = \sum_{j=1}^M \lP_j(0)
\]
the projector onto the eigenspace $\cE_0 = \mathrm{Ran}(\cP_0)$
spanned by the $N$  degenerate eigenstates of $H_0$.
For simplicity, we assume
that the perturbation $V$ is sufficient to split the degeneracy (so that $M = N$), in
the sense that the following assumption holds true.

\begin{assumption}[Degeneracy splitting]
\label{ass:gap}
The finite rank self-adjoint operator $\cP_0 V \cP_0 \, : \, \cE_0 \to \cE_0$ has
non-degenerate eigenvalues, and
there is a gap between the $N$ first levels in the interval $(0,1]$:
for any $\lambda^\ast > 0$, there exists $\alpha$ (depending on $\lambda^\ast$) such that
\begin{equation}
  \label{eq:local_gap}
  \inf_{\lambda^\ast \leq \lambda \leq 1}
  \min_{k \not = l}
  \left| \lE_k(\lambda) - \lE_l(\lambda) \right| \geq \alpha > 0.
\end{equation}
\end{assumption}

This implies that the projectors $\lP_j(\lambda)$ are rank-1 projectors for any
$\lambda > 0$ (since it can be proved that the perturbation $V$ is enough to split
the eigensubspaces, and the gap condition on $(0,1]$ ensures that no crossing
can happen; see Section~\ref{sec:geometric} for more details).

\begin{remark}\label{Rem Gap condition}
  Assumption~\ref{ass:gap} may be relaxed in several ways.
  First, the operator $\cP_0 V \cP_0$ can have degenerate eigenvalues,
  but then higher order terms should be considered in the perturbative
  expansion of the eigenvalues. The gap
  assumption can be relaxed as well, and some crossings could be
  allowed.
    Besides, the general case of $M < N$ projectors of ranks greater or
    equal to 1 can be treated similarly upon modifying the condition
    $\left \| \lP_j(1) - \lP_j(0) \right \| < 1$ required in
    Theorem~\ref{thm:GLM} below. All these extensions are discussed in
    Section~\ref{sec:extension}.
\end{remark}

\subsection{Switching procedure}

Consider, for $\tau \in (-\infty,0]$,
\[
H(\tau) = \lH(f(\tau)) = H_0 + f(\tau) \, V,
\]
where the switching function $f$ has values in $[0,1]$
(in order for the operator $H(\tau)$ to be well-defined as a self-adjoint operator
on $D(H_0)$).
We denote by $P_j(\tau)$ the eigenprojectors and eigenvalues
corresponding to the first $N$ eigenvalues $E_j(\tau)$ of $H(\tau)$;
also, $P_{N+1}(\tau) = \1 - \sum_{k=1}^N P_k(\tau)$. Of course,
\[
P_j(\tau) = \lP_j(f(\tau)), \qquad E_j(\tau) = \lE_j(f(\tau)).
\]

\goodbreak

For the subsequent analysis, we assume that

\begin{assumption}
  \label{ass:f_adiabatic}
  The switching function $f \, : \, (-\infty,0] \to [0,1]$
  is a $\mathrm{C}^2$ function such that
  \begin{enumerate}[(i)]
  \item $f$ is non-decreasing;
  \item $f(0) = 1$ and $\dps \lim_{\tau \to -\infty} f(\tau) = 0$;
  \item $f,f'' \in \mathrm{L}^1((-\infty,0])$.
  \end{enumerate}
\end{assumption}

The most common choice in practice is $f(\tau) = \rme^\tau$. Notice
however that any $\mathrm{C}^2$ non-decreasing compactly supported
function with $f(0)=1$ satisfies the above assumptions. In the latter case, the
monotonicity of $f$ implies that the support of $f$ is a compact
interval $[R_f,0]$, and $f(t) > 0$ for $t \in (R_f,0]$. The assumption
$f \in \mathrm{C}^2$ ensures that the
adiabatic evolution (see~\eqref{eq:adiabatic_unitary_evolution}
below) is well-defined.

As a consequence of these assumptions, $f' \geq 0$
and $f' \in \mathrm{L}^1((-\infty,0]) \cap
\mathrm{L}^\infty((-\infty,0])$, hence $f' \in
\mathrm{L}^2((-\infty,0])$. Indeed, the boundedness of $f'$ is a
consequence of the fundamental theorem of calculus and the fact
that $f'' \in \mathrm{L}^1((-\infty,0])$. Besides, $ \int_t^0 f' = f(0) -
f(t) \leq 1, $ and $f' \geq 0$, hence $f' \in
\mathrm{L}^1((-\infty,0])$.

\begin{remark}
It can be shown that eigenprojectors and eigenvectors are analytic
with respect to $\lambda = f(\tau)$ (see Section~\ref{sec:geometric}).
When the switching function~$f$
is analytic, the eigenvalues $E_j(\tau)$ (and the associated eigenvectors and eigenprojectors)
are also analytic with respect to $\tau$.
\end{remark}

We denote by $U_\eps(s,s_0)$ the unitary evolution  generated by
$H(\eps s)$, \ie the unique solution (which is well-defined
by Theorem~X.70 in~\cite{ReedSimon2}) of the problem:
\[
\ri \frac{dU_\eps(s,s_0)}{ds} = H(\eps s) \, U_\eps(s,s_0), \quad
U_\eps(s_0,s_0) = \1.
\]
In order to remove divergent phase factors (see the proof in
Section~\ref{sec:adiabatic_switching}), it is convenient to consider
evolution operators in the interaction picture:
\[
U_{\eps,\rm int}(s,s_0) = \rme^{\ri s H_0} U_\eps(s,s_0) \, \rme^{-\ri s_0 H_0}.
\]
It is actually more convenient to rescale the time and to consider a
macroscopic time $t = \eps s$. The unitary evolution $U^\eps(t,t_0)$ in terms of the
macroscopic time is the solution of
\[
\ri \eps \frac{dU^\eps(t,t_0)}{dt} = H(t) U^\eps(t,t_0), \quad
U^\eps(t_0,t_0) = \1,
\]
and, in the interaction picture,
\[
U_{\rm int}^\eps(t,t_0) = \rme^{\ri t H_0/\eps} \, U^\eps(t,t_0) \, \rme^{-\ri t_0 H_0/\eps}.
\]
Standard results show that $U_{\rm int}^\eps(t,-\infty) \psi = \lim_{t_0 \to -\infty}
U_{\rm int}^\eps(t,t_0) \psi$ exists for $\psi \in D(H_0)$ (for instance, by using a
Cook's type argument and rewriting
this operator as the integral of its derivative with respect to $t_0$).

\subsection{Main results}

We are now in position to state our main results.

\begin{theoreme}
\label{thm:GLM} Suppose that the gap conditions on~$H_0$ and~$V$
(Assumptions~\ref{ass:structure} and~\ref{ass:global_gap}) are
satisfied, and that the perturbation term $V$ lifts the degeneracy
(Assumption~\ref{ass:gap}). Consider a switching function verifying
Assumption~\ref{ass:f_adiabatic}. Let $(\psi_1,\dots,\psi_N)$ be an
orthonormal basis of $\cE_0$ which diagonalizes the bounded operator $\cP_0 V
\cP_0 \big |_{\cE_0}$. Then, if
\begin{equation}
  \label{eq:condition_for_existence_ratio}
  \left \| P_j(-\infty) - P_j(0) \right \| < 1,
\end{equation}
the limit
\begin{equation}\label{Eq: GL limit}
\Psi_j = \lim_{\eps \to 0} \frac{U_{\rm int}^\eps(0,-\infty) \psi_j}
    {\bra \psi_j \rsep \left. U_{\rm int}^\eps(0,-\infty) \psi_j \ket}
\end{equation}
exists and is an eigenstate of $H_0 + V$.
\end{theoreme}

Notice that, for a generic state $\psi \in \Ran \cP_0$ which is not
an eigenvector of $\cP_0 V \cP_0 \big |_{\cE_0}$ the above limit
 generically does not exist, as showed in
\cite{BSP08} by using a simple toy model. It is therefore crucial to
select the appropriate initial states, so that the Gell-Mann \& Low
limit (\ref{Eq: GL limit}) does exist.

\bigskip

As an intermediate step, the eigenprojector $P_j(0)$ and a
corresponding eigenfunction $\Psi_j$ can be recovered by Kato's
geometric evolution \cite{Kato50}.

\begin{definition}\label{def Kato evolution}
The Kato evolution operator $A(s,s_0)$, for $s,s_0\in \R$ is the
unique solution of the problem
\begin{equation}
  \label{eq:intertwiner_s}
  \frac{dA(s,s_0)}{ds} = K(s) \, A(s,s_0),
  \quad
  A(s_0,s_0) = \1,
\end{equation}
with
\[
K(s) = - \sum_{j=1}^{N+1} P_j(s) \, \frac{dP_j}{ds}(s).
\]
\end{definition}

\noindent By our assumptions, the operator $K(s)$ is uniformly
bounded (see the comment after Definition~\ref{Def:adiabatic
evolution}). The Kato evolution operator is a unitary operator which
intertwines the spectral subspaces of $H(s)$ and $H(s_0)$, in the
sense that
\[
A(s, s_0) P_j(s_0) = P_j(s) A(s,s_0).
\]

Equipped with this notation, we have the following result,
where no condition analogous to
(\ref{eq:condition_for_existence_ratio}) is assumed.

\begin{proposition}
\label{Prop:geometric evolution} Let
Assumptions~\ref{ass:structure}, \ref{ass:global_gap}, \ref{ass:gap}
and~\ref{ass:f_adiabatic} be satisfied.
Let $(\psi_1,\dots,\psi_N)$ be an orthonormal basis of $\cE_0$ which
diagonalizes the operator $\cP_0 V \cP_0 \big |_{\cE_0}$. Then
\[
\Psi_j := A(0, -\infty) \psi_j
\]
is an eigenvector of $H_0 + V$.
\end{proposition}

It is actually much simpler to consider the geometric evolution operator $A$ rather
than the evolution operator $U^\eps_{\rm int}$ since less conditions on $H_0$ and $V$ are
required. Indeed, there is no denominator which needs to be considered in
order to remove a divergent phase.
However, the many-body theory used in physics is
defined in terms of $U_{\rm int}^\varepsilon$ and not in terms of~$A$.

\bigskip

We sketch shortly the structure of the proof, which is done in four
steps:
\begin{enumerate}[(i)]
\item first, we use the Kato geometric evolution backward in time, in
  order to identify, though in a non explicit manner, the initial
  subspaces of $\cP_0$ whose vectors can be considered as convenient
  initial states;
\item in a second step
  (Section~\ref{sec:characterization}), we give an explicit
  description of these initial subspaces, in terms of the eigenvectors
  of $\cP_0 V \cP_0 \big |_{\cE_0}$.
  At this stage, we are already in position to prove
  Proposition~\ref{Prop:geometric evolution};
\item then, we show how the limit of
  the full evolution $U^\eps_{\rm int}$ can be related to the
  geometric evolution as $\eps \to 0$
  (Section~\ref{sec:full_evolution_limit}). A first step is to introduce
  an intermediate concept, the adiabatic evolution, which
  takes some dynamics into account (arising from the Hamiltonian
  operator). The adiabatic evolution is also an intertwiner. Since
  intertwiners differ only by a phase (in a sense to be made precise),
  and, provided this phase can be removed, the adiabatic evolution can
  be reduced to the geometric one (see Section~\ref{sec:adiabatic_switching});
\item the last point is to show that
  the limit as $\eps \to 0$ of the full evolution is the adiabatic
  evolution (see Section~\ref{sec:adiabatic_limit}).
\end{enumerate}
Steps~(iii) and~(iv) are straightforward extensions of previous results
in adiabatic theory, and we heavily relied on the paper by Nenciu
and Rasche~\cite{NR89} for Section~\ref{sec:adiabatic_switching} and
the book by Teufel~\cite{Teufel} for Section~\ref{sec:adiabatic_limit}.

%
%

\section{Proof of the results}
\label{sec:proof}

\subsection{Geometric evolution and definition of the initial states}
\label{sec:geometric}

In view of the local gap assumption, the projectors and eigenvalues
of $\lH(\lambda)$ are real analytic functions of $\lambda \in
(0,1]$. Besides, Theorem II.6.1 in~\cite{Kato} shows that the
eigenvalues $\lE_j$ and projectors $\lP_j$ can be analytically
continued in the limit $\lambda \to 0$. The Kato construction
of unitary operators $A$ intertwining projectors can then be
performed, see for instance Theorem~XII.12 in \cite{ReedSimon4} or
Sections~II.4 and~II.6.2 in~\cite{Kato}. Consider the operator
\[
\lK(\lambda) = - \sum_{j=1}^{N+1} \lP_j(\lambda) \, \frac{d\lP_j}{d\lambda}(\lambda),
\]
first proposed in~\cite{Kato50}, and the unique solution of
\begin{equation}
  \label{eq:Kato_Nagy}
  \frac{d\lA(\lambda,\lambda_0)}{d\lambda} = \lK(\lambda) \, \lA(\lambda,\lambda_0),
  \quad
  \lA(\lambda_0,\lambda_0) = \id.
\end{equation}
Since $\lK(\lambda)$ is uniformly bounded, the operator
$\lA(\lambda,\lambda_0)$ is well-defined and strongly continuous
(see Theorem~X.69 in~\cite{ReedSimon2}). Besides,
$\lA(\lambda,\lambda_0)$ is unitary, and intertwines the spectral
subspaces:
\[
\lP_j(\lambda) = \lA(\lambda,\lambda_0) \lP_j(\lambda_0) \lA(\lambda,\lambda_0)^\ast.
\]
It is also easily shown that
$\lA(\lambda_2,\lambda_1) \lA(\lambda_1,\lambda_0) = \lA(\lambda_2,\lambda_0)$, for instance
by computing the derivative of both expressions with respect to $\lambda_1$ and
using the uniqueness of the solution of~\eqref{eq:Kato_Nagy}.

\bigskip

We define the \emph{initial subspaces} by evolving backwards
eigenstates of the Hamiltonian $\lH(\lambda)$ for which the
perturbation has split the degeneracy: the corresponding
eigenprojector is defined as
\begin{equation}
  \label{eq:initial_states_geometric}
  P^{\rm init}_j := \lA(0,\lambda) \lP_j(\lambda) \lA(\lambda,0),
\end{equation}
the definition being independent of $\lambda > 0$.

Eigenstates of $\lH(1) = H_0 + V$ are then obtained by evolving
initial states belonging to the range of $P_j^{\rm init}$ according
to the Kato evolution operator. Indeed, $\lA(1,0) P_j^{\rm init} =
\lA(1,0) \lA(0,\lambda) \lP_j(\lambda) = \lA(1,\lambda)
\lP_j(\lambda)$. Thanks to the intertwining property of $A$, it
holds
\begin{equation}
  \label{Eq: initial eigenspace}
  \lP_j(1) = \lA(1,0) P_j^{\rm init} \lA(0,1).
\end{equation}

\subsection{Characterization of the initial states}
\label{sec:characterization}

The above paragraph shows that it is crucial to identify 
$\mathrm{Ran}(P_j^{\rm init})$. We now \emph{characterize} these spaces by an explicit condition.

\paragraph{General expressions of the eigenvalues and eigenvectors.}
Since the eigenvalues and eigenprojectors of $\lH(\lambda)$ are
analytic in $\lambda \in [0,1]$, the following expansions are
valid for $1 \leq j \leq N$:
\begin{equation}
  \label{eq:expansion_E}
  \lE_j(\lambda) = \sum_{n=0}^{+\infty} \lambda^n E_{j,n},
\end{equation}
and
\[
\lP_j(\lambda) = \sum_{n=0}^{+\infty} \lambda^n P_{j,n}.
\]
Of course, $E_{j,0} = E_0 = \lE_j(0)$, the common value of the
energy in the degenerate ground-state. Notice also that the operators
$P_{j,n}$ are not necessarily orthogonal projectors.

To define $\lP_j(\lambda)$, it is more convenient to consider an eigenvector $\phi_j(\lambda)$
associated with $\lE_j(\lambda)$, \ie a non-zero element
of $\mathcal{H}$ satisfying
\begin{equation}
  \label{eq:eigenequation}
  \lH(\lambda) \phi_j(\lambda) = \lE_j(\lambda) \, \phi_j(\lambda).
\end{equation}
Such an eigenvector can be chosen to be analytic, by the same
results which allow to conclude to the analyticity of the
eigenprojectors. We therefore write
\begin{equation}
  \label{eq:expansion_phi}
  \phi_j(\lambda) = \sum_{n=0}^{+\infty} \lambda^n \varphi_{j,n}.
\end{equation}
Once such an eigenvector is known, the
analytic eigenprojector can be constructed as
\[
\lP_j(\lambda) = \lsep \frac{\phi_j(\lambda)}{\|\phi_j(\lambda)\|} \ket
\bra \frac{\phi_j(\lambda)}{\|\phi_j(\lambda)\|} \rsep.
\]

The aim of this section is to provide
an explicit expression of the leading terms of the above expansions,
in order to have a more explicit definition of $P_j^{\rm init}$.
To this end, we first construct a basis of $\cE_0$, which will turn out
to be particularly useful to characterize the terms in the expansions
\eqref{eq:expansion_E} and \eqref{eq:expansion_phi}.

\paragraph{Diagonalization of $\cP_0 V \cP_0$.}
Since $\cP_0 V \cP_0$ and $\cP_0$ commute,
it is possible to construct an orthonormal basis
$(\varphi_{1,0},\dots,\varphi_{N,0})$ of $\cE_0$ such that
\begin{equation}
  \label{eq:PVP_ortho}
  \cP_0 V \cP_0 \, \varphi_{j,0} = \alpha_j \varphi_{j,0}
\end{equation}
for some real numbers $\alpha_1,\dots,\alpha_N$, and
\begin{equation}
  \label{eq:PVP_basis}
  \forall j \neq k, \quad
  \bra \varphi_{k,0}, \, \cP_0 V \cP_0 \, \varphi_{j,0} \ket = 0.
\end{equation}



\paragraph{Expressions for the terms in the expansions
  \eqref{eq:expansion_E}-\eqref{eq:expansion_phi} at order 1.}
We identify the terms associated with the same powers of $\lambda$
in \eqref{eq:eigenequation}. An additional normalization condition
should be added in order to uniquely define the solution, so we
impose
\begin{equation}
  \label{eq:normalization}
  \forall \lambda \in [0,1],
  \qquad
  \bra \varphi_{j,0}, \, \phi_j(\lambda) \ket = 1,
\end{equation}
as is done in~\cite{RefGianluca}. As will be seen below, this
condition is simpler to work with than the standard condition
$\|\phi_j(\lambda)\|=1$. The identification of the terms in
\eqref{eq:eigenequation} gives, for $1 \leq j \leq N$, the following
hierarchy of equations:

\begin{eqnarray*}
  (H_0 - E_0) \varphi_{j,0}   &=& 0,\\
  (H_0 - E_0) \varphi_{j,1} &=& (E_{j,1}-V) \varphi_{j,0},           \\
  (H_0 - E_0) \varphi_{j,2} &=& (E_{j,1}-V) \varphi_{j,1} +
  E_{j,2}\varphi_{j,0},
\end{eqnarray*}
and, for $n \geq 2$,
\begin{equation}
  \label{eq:eq_ordre_n}
  (H_0 - E_0) \varphi_{j,n+1} = (E_{j,1}-V) \varphi_{j,n} + \sum_{m=0}^{n-1}
  E_{j,n+1-m} \varphi_{j,m}.
\end{equation}
The equation on the term of order zero does not give any
information on the choice of the initial states $\varphi_{j,0}$. This
information can be obtained from the first order condition:
\begin{equation}
  \label{eq:eq_ordre_1}
  (H_0 - E_0) \varphi_{j,1} = (E_{j,1}-V) \varphi_{j,0}.
\end{equation}
A necessary condition for this equation to have a solution is that
the right-hand side belongs to $\cE_0^\perp$ (since the
left-hand side does):
\begin{equation}
  \label{eq:conditions_on_the_basis}
  \forall 1 \leq j,k \leq N, \quad \bra \varphi_{k,0}, (E_{j,1}-V) \varphi_{j,0} \ket = 0.
\end{equation}
This requires
\[
E_{j,1} =  \bra \varphi_{j,0}, V \varphi_{j,0} \ket,
\]
and
\[
\forall k \not = j, \quad \bra \varphi_{k,0}, V \varphi_{j,0} \ket = 0.
\]
Therefore, the conditions \eqref{eq:conditions_on_the_basis} for $k
\not = j$ cannot be fulfilled for a general basis. A necessary
condition is that the basis $\{\varphi_{k,0}\}_{k=1,\dots,N}$
of $\cE_0$ diagonalizes $\cP_0 V \cP_0$. Besides,
the first-order term in the energy shifts are exactly the
eigenvalues of $\cP_0 V \cP_0$.
This condition determines uniquely the basis when $\cP_0 V \cP_0$
has non-degenerate eigenvalues. If this is not the case, information about the higher
order equations in the hierarchy is needed (see Section~\ref{sec:extension}).

\begin{remark}
Assuming that the bands do not recross after the initial splitting,
and if the degenerate state is the ground state of $H_0$, then the ground state of
$H_0 + V$ is obtained by following the eigenstate
associated with the lowest $E_{j,1}$.
\end{remark}

Once the initial basis and the first energy shifts have been defined, the first
order term in the variation of the eigenstates can be obtained
from \eqref{eq:eq_ordre_1} as the sum of the reduced resolvent applied to the right hand
side, and some solution of the homogeneous equation $(H_0-E_0) \psi = 0$:
\beqn
\label{eq:1st_order_phi}
\varphi_{j,1} & = & \sum_{k=1}^N c^1_{j,k} \varphi_{k,0} +
\left ( H_0 - E_0 \right )^{-1} \Big |_{\cE_0^\perp} (E_{j,1}-V) \varphi_{j,0}
\\
& = &  \sum_{k \not = j} c^1_{j,k} \varphi_{k,0}
-R_0 V \varphi_{j,0},
\eeqn
where
\[
R_0 = ( H_0 - E_0 )^{-1} \Big |_{\cE_0^\perp}
= (\id - \cP_0) \left( H_0 - E_0 \right)^{-1} (\id - \cP_0)
\]
is a bounded operator from $\cE_0^\perp$ to $\cE_0^\perp \cap D(H_0)$, and
$c^1_{j,j} = 0$ in view of the normalization
condition~\eqref{eq:normalization}. The coefficients $c^1_{k,j}$
(for $k \neq j$) are undetermined at this stage. They have to be
chosen so that the right hand side of the next equation in the hierarchy
is in $\cE_0^\perp$.

\paragraph{Conclusion: characterization of the initial subspaces.}
The above computations show that $\lP_j(\lambda) = P_{j,0} + \OO(\lambda)$,
with $P_{j,0} = |\varphi_{j,0} \rangle \langle \varphi_{j,0}|$.
Besides, $\|\lA(0,\lambda) - \id \| = \OO(\lambda)$ in view of the differential
equation~\eqref{eq:Kato_Nagy} satisfied by $\lA$.
The initial subspace \eqref{eq:initial_states_geometric} is therefore
\[
P_j^{\rm init} = \lA(0,\lambda) P_j(\lambda) =
\lim_{\lambda \to 0} \lA(0,\lambda)\left [ P_{j,0} + \OO(\lambda) \right]
= P_{j,0}.
\]

\paragraph{Proof of Proposition \ref{Prop:geometric evolution}.}
Let $\psi \in \cE_0$ be an eigenvector of $\cP_0 V \cP_0$. Then, there exists 
$j \in \{ 1, \dots,N \}$ such that $\psi \in \Ran(P_{j,0})=\Ran(P_j^{\rm init})$. Using
(\ref{Eq: initial eigenspace}), it follows
\[
A(0, -\infty)\psi_j = \tilde A(1,0) \psi_j \in \Ran\left(\widetilde{P}_j(1)\right),
\]
which proves the claim.

\subsection{Adiabatic evolution and limit of the full evolution}
\label{sec:full_evolution_limit}

\begin{definition}\label{Def:adiabatic evolution}
The adiabatic evolution operator $\UA(s, s_0)$ is defined for
$(s,s_0) \in \R^2$ as the unique solution of the problem
\begin{equation}
  \label{eq:adiabatic_unitary_evolution}
  \ri \frac{d\UA(s,s_0)}{ds} = \HA(s) \UA(s,s_0),
  \quad
  \UA(s_0,s_0) = \id,
\end{equation}
where the adiabatic Hamiltonian is
\[
\HA(s) = H(s) + \ri K(s),
\]
with
\[
K(s) = - \sum_{j=1}^{N+1} P_j(s) \, \frac{dP_j}{ds}(s).
\]
\end{definition}

\noindent
Notice that $K(s) = f'(s) \, \widetilde{K}(f(s))$ so that
\begin{equation}
  \label{eq:boundedness_of_K}
  \| K(s) \| \leq C f'(s)
\end{equation}
for some constant $C > 0$. Therefore, $K(s)$ is uniformly bounded since $f'$ is bounded
by our assumptions on the switching function.

Compared to the geometric evolution~\eqref{eq:Kato_Nagy}, a
Hamiltonian term has been added, which is at the origin of some
dynamical phase factor in the dynamics. The adiabatic dynamics is
well-defined in view of the assumptions made on $H_0$, $V$
and~$f$ (see Theorem~X.70 in~\cite{ReedSimon2}). A simple computation shows that it
intertwines the spectral subspaces:
\[
P_j(s) = \UA(s,s_0) P_j(s_0) \UA(s,s_0)^\ast.
\]
Switching to the interaction picture, we define
\[
\UAI(s,s_0) = \rme^{\ri s H_0} \, \UA(s,s_0) \, \rme^{-\ri s_0 H_0}.
\]
The factor $\eps$ is introduced by slowing down the switching as
\begin{equation}
  \label{eq:adiabatic_unitary_evolution_eps}
  \ri \frac{d\UAe(s,s_0)}{ds} = \HA(\eps s) \UAe(s,s_0),
  \quad
  \UAe(s_0,s_0) = \id,
\end{equation}
and the corresponding operator in the interaction picture is
$\rme^{\ri s H_0} \UAe(s,s_0) \, \rme^{-\ri s_0 H_0}$.
It is convenient to rewrite the evolution \eqref{eq:adiabatic_unitary_evolution_eps}
in the rescaled time variable $t = \eps s$:
\begin{equation}
  \label{eq:adiabatic_unitary_evolution_rescaled}
  \ri \eps \frac{d\UA^\eps(t,t_0)}{dt} = \HA^\eps(t) \UA^\eps(t,t_0),
  \quad
  \UA^\eps(t_0,t_0) = \id,
\end{equation}
with $\HA^\eps(t) = H(t) + \ri \eps K(t)$.
The associated operator in the interaction
picture is
\[
\UAI^\eps(t,t_0) = \rme^{\ri t H_0/\eps} \UA^\eps(t,t_0) \, \rme^{-\ri t_0 H_0/\eps}.
\]
Theorem~\ref{thm:GLM} is then a consequence of the following results.

\begin{lemma}
  \label{Lemma:adiabatic evolution}
  Let $\psi_j \in P_j^{\rm init}$
  (defined by \eqref{eq:initial_states_geometric}). Then, under the
  assumptions of Theorem~\ref{thm:GLM}, the vector
  \begin{equation}
    \label{eq:equality_ratio}
    \frac{\UAI^\eps(0,-\infty) \psi_j}{\bra \psi_j \rsep \UAI^\eps(0,-\infty) \psi_j \rangle}
    =
    \frac{\UAI(0,-\infty) \psi_j}{\bra \psi_j \rsep \UAI(0,-\infty) \psi_j \rangle}
  \end{equation}
  is an eigenstate of $H_0$.
\end{lemma}

\begin{lemma}
  \label{Lemma:adiabatic limit}
  Let $\psi_j \in P_j^{\rm init}$. Then, under the
  assumptions of Theorem~\ref{thm:GLM},
  \[
  \lim_{\eps \to 0} \left ( \frac{U_{\rm int}^\eps(0,-\infty)
    \psi_j}{\bra \psi_j \rsep
    U_{\rm int}^\eps(0,-\infty) \psi_j \rangle}
  - \frac{\UAI^\eps(0,-\infty) \psi_j}{\bra \psi_j \rsep
    \UAI^\eps(0,-\infty) \psi_j \rangle} \right ) = 0.
  \]
\end{lemma}

\subsubsection{Proof of Lemma \ref{Lemma:adiabatic evolution}}
\label{sec:adiabatic_switching}

We show first in this section that $\psi_j$ can be transformed into
an eigenstate of $H(0) = \lH(1)$ using the adiabatic evolution
defined from \eqref{eq:adiabatic_unitary_evolution}, and then the
equality of the ratios \eqref{eq:equality_ratio}. The proof
presented here reproduces the argument of Nenciu and
Rasche~\cite{NR89}, which was given in the case $N=1$ with our
notation, but can be applied {\it mutatis mutandis} to the case
considered here. We however present the proof for completeness.

\paragraph{Evolution in the case $\eps = 1$.}

Since both $\UA$ and $A$ are intertwiners, they differ only by a phase which commutes with
the spectral projectors. Indeed, define
\[
\Phi(s,s_0) = A(s,s_0)^\ast \UA(s,s_0),
\]
so that $\UA(s,s_0) = A(s,s_0) \, \Phi(s,s_0)$.
Then,
\[
[\Phi(s,s_0),P_j(s_0)] = 0,
\]
as can be seen using the intertwining properties:
\begin{eqnarray*}
[\Phi(s,s_0),P_j(s_0)] & = & A(s,s_0)^\ast \UA(s,s_0) P_j(s_0) - P_j(s_0) A(s,s_0)^\ast \UA(s,s_0) \\
& = & A(s,s_0)^\ast P_j(s) \UA(s,s_0) - A(s,s_0)^\ast P_j(s) \UA(s,s_0) = 0.\\
\end{eqnarray*}
The time-evolution of the phase matrix can be simplified due to this commutation property.
First,
\[
\frac{d\Phi(s,s_0)}{ds} = -\ri A(s,s_0)^\ast H(s) \UA(s,s_0),
\]
since $K(s)^\ast = -K(s)$.
Besides,
\[
\Phi(s,s_0) = \left ( \sum_{k = 1}^{N+1} P_k(s_0) \right ) \Phi(s,s_0)
\left ( \sum_{k = 1}^{N+1} P_k(s_0) \right ) = \sum_{k=1}^{N+1} \Phi_k(s,s_0),
\]
where $\Phi_k(s,s_0) = P_k(s_0) \Phi(s,s_0) P_k(s_0)$.
The time evolution of the projected phase-matrix is a scalar phase since
\[
\frac{d}{ds} \Phi_k(s,s_0) = -\ri E_k(s) \Phi_k(s,s_0),
\]
hence
\[
\Phi(s,s_0) P_j(s_0) = \exp\left( -\ri \int_{s_0}^s E_j(r) \, dr \right) P_j(s_0).
\]
The geometric evolution and the adiabatic evolution are therefore
related through some global dynamical phase:
\[
\UA(s,s_0) P_j(s_0) = A(s,s_0) \Phi(s,s_0) P_j(s_0)
= \exp\left( -\ri \int_{s_0}^s E_j(r) \, dr \right) A(s,s_0) P_j(s_0).
\]

To describe the asymptotic evolution, we follow closely the approach of
\cite{NR89}. In order for $\UA(s,s_0) P_j(s_0)$ to be defined in the
limit $s_0 \to -\infty$, it is important to work in the interaction
picture. Then,
\begin{eqnarray*}
\UAI(s,s_0) P_j(-\infty)
& = & \rme^{\ri s H_0} A(s,s_0) \Phi(s,s_0) \rme^{-\ri s_0 H_0} P_j(-\infty)\\
& = & \rme^{-\ri s_0 E_0} \rme^{\ri s H_0} A(s,s_0) \rme^{-\ri s H_0}
\rme^{\ri s H_0} \Phi(s,s_0) P_j(-\infty). \\
\end{eqnarray*}
Using
\[
\Phi(s,s_0) P_j(s_0) =
P_j(s_0) \Phi(s,s_0) P_j(s_0) = \exp \left (-\ri\int_{s_0}^s E_j(r) \, dr \right ) P_j(s_0),
\]
it holds
\begin{eqnarray*}
& & \hspace{-1cm} \rme^{-\ri s_0 E_0} \rme^{\ri s H_0} \Phi(s,s_0) P_j(-\infty) \\
& = & \rme^{-\ri s_0 E_0} \rme^{\ri s H_0} \Phi(s,s_0) P_j(s_0)
+ \rme^{-\ri s_0 E_0} \rme^{\ri s H_0} \Phi(s,s_0) (P_j(-\infty)-P_j(s_0)) \\
& = & \exp \left (-\ri \int_{s_0}^s E_j(r) \, dr - \ri s_0 E_0 \right )
\rme^{\ri s H_0} P_j(s_0)
+ \rme^{-\ri s_0 E_0} \rme^{\ri s H_0} \Phi(s,s_0) (P_j(-\infty)-P_j(s_0)) \\
& = & \exp \left (-\ri \int_{s_0}^s E_j(r) \, dr - \ri s_0 E_0 \right )
\left [ \rme^{\ri s H_0} P_j(-\infty)
+ \rme^{\ri s H_0} (P_j(s_0)-P_j(-\infty)) \right ] \\
& & \quad + \rme^{-\ri s_0 E_0} \rme^{\ri s H_0} \Phi(s,s_0) (P_j(-\infty)-P_j(s_0)) \\
& = & \exp \left (-\ri \int_{s_0}^s E_j(r)-E_0 \, dr\right ) P_j(-\infty)
+ W(s,s_0) (P_j(s_0)-P_j(-\infty)), \\
\end{eqnarray*}
where $\| W \| \leq 2$.
Since $\lambda \mapsto E_j(\lambda)$ is $C^1$ on the
compact interval $[0,1]$, there exists a constant $C > 0$ such that
\[
\left|E_j(r)-E_0\right| = \left|\lE_j(f(r)) - \lE_j(0)\right| \leq C f(r).
\]
Since $f \in {\rm L}^1((-\infty,0])$, this shows that the function
$r \mapsto E_j(r)-E_0$ is integrable on $(-\infty,0]$.
Besides, $P(s_0) \to P_j(-\infty)$ when $s_0 \to -\infty$.
The limit $s_0 \to -\infty$ of
$\UAI(s,s_0) P_j(-\infty)$ is therefore well-defined:
\begin{equation}
  \label{eq:UAI_s}
  \UAI(s,-\infty) P_j(-\infty) =
  \exp \left (-\ri \int_{-\infty}^s E_j(r)-E_0 \, dr\right )
  \rme^{\ri s H_0} A(s,-\infty) \rme^{-\ri s H_0} P_j(-\infty).
\end{equation}
The above equality reads, for $s=0$,
\[
\UAI(0,-\infty) P_j(-\infty) = \exp \left (-\ri \int_{-\infty}^0 E_j(r)-E_0 \, dr\right )
A(0,-\infty) P_j(-\infty).
\]
Since $P_j(0) A(0,-\infty) = A(0,-\infty) P_j(-\infty)$,
it holds, for $\psi_j \in P_j^{\rm init} = P_j(-\infty) =
\mathrm{Ran}(\varphi_{j,0})$,
\begin{equation}
  \label{eq:1D_miracle}
  P_j(0) \psi_j = A(0,-\infty) P_j(-\infty) A(0,-\infty)^\ast \psi_j
  = \bra \psi_j \rsep \left. A(0,-\infty)^\ast \psi_j \ket \, A(0,-\infty) \psi_j.
\end{equation}
Finally,
\[
\frac{P_j(0) \psi_j}{\| P_j(0) \psi_j \|^2}
= \frac{P_j(0) \psi_j}{\bra \psi_j \rsep \left. P_j(0) \psi_j \ket}
= \frac{A(0,-\infty) \psi_j}{\bra \psi_j \rsep \left. A(0,-\infty) \psi_j \ket}
= \frac{\UAI(0,-\infty) \psi_j}{\bra \psi_j \rsep \left. \UAI(0,-\infty) \psi_j \ket},
\]
which shows that the adiabatic evolution transforms the initial eigenstate into an
eigenstate of $H(1)$ provided $\| P_j(0) \psi_j \| \not = 0$, which is the case
when $\| P_j(0) - P_j(-\infty) \| < 1$.

\paragraph{Evolution in the case $\eps > 0$.}
Let us conclude this section by proving the equality \eqref{eq:equality_ratio}.
Computations similar to the ones performed in the case $\varepsilon = 1$ lead to
\[
\UAI^\eps(0,-\infty) P_j(-\infty) =
\exp \left (-\frac{\ri}{\eps} \int_{-\infty}^0 E_j(\tau)-E_0 \, d\tau\right )
A(0,-\infty) P_j(-\infty).
\]
This can be seen for instance by noticing that
\eqref{eq:adiabatic_unitary_evolution_rescaled} can be rewritten in
the form \eqref{eq:adiabatic_unitary_evolution}, upon considering
the Hamiltonian $H/\eps$ in~\eqref{eq:adiabatic_unitary_evolution}. 
Therefore, $\UAI^\eps(0,-\infty)
P_j(-\infty)$ is equal, up to the $\eps$-dependence in the
phase factor, to $\UAI(0,-\infty) P_j(-\infty)$.
The non convergent phase factor can be eliminated precisely by considering
the Gell-Mann and Low ratio~\eqref{eq:equality_ratio}.

\subsubsection{Proof of Lemma \ref{Lemma:adiabatic limit}}
\label{sec:adiabatic_limit}


It is sufficient to prove that
\[
\lim_{\eps \to 0} \| U^\eps(0,-\infty) - \UA^\eps(0,-\infty) \| = 0,
\]
which indeed gives the result since
\[
\| U_{\rm int}^\eps(t,t_0) - \UAI^\eps(t,t_0) \| =
\| U^\eps(t,t_0) - \UA^\eps(t,t_0) \|.
\]
Notice that, although none of the operators
$U^\eps(0,-\infty), \UA^\eps(0,-\infty)$ has a limit when $\eps \to 0$, the
difference goes to 0 in this limit.

\bigskip

The proof is based on the proofs of Theorem~2.2 and Corollary~2.5
in the book by Teufel~\cite{Teufel},
which are extended to the case of non-compactly supported switching functions
and $N > 1$ with our notation.
In this section, $C$ and $C'$ denote constants, which may change from line to line,
but are always independent of $t$, $\eps$, etc, and depends only on the relative
$H_0$-bound of $V$, on $N$, on $\Delta^\ast$ and on bounds on the functions $\lP_j$ and their
derivatives on $[0,1]$.

We denote by $\delta_j(t) \geq 0$ the local gap around $E_j(t)$:
\[
\delta_j(t) = \min \Big \{ |E_j(t)-E|, \
E \in \sigma(H(t)) \backslash \{ E_j(t) \} \Big \}.
\]
Notice that $\delta_j(t) > 0$ when $f(t) > 0$, but $\delta_j(t) \to 0$ when
$f(t) \to 0$ since the initial
eigenvalue is $N$-fold degenerate (see Assumption~\ref{ass:gap}).
In fact, the analysis of Section~\ref{sec:characterization} shows that
there exist $\alpha_1,\alpha_2 > 0$ such that
\begin{equation}
  \label{eq:local_gap_estimate}
  \alpha_1 \leq \left | \frac{\delta_j(t)}{f(t)} \right | \leq \alpha_2
\end{equation}
when $f(t) > 0$.

\paragraph{Rewriting the difference as an integral.}
The difference between the two unitary evolutions is rewritten as the integral
of the derivative, as:
\begin{eqnarray*}
U^\eps(t,t_0) - \UA^\eps(t,t_0) & = & -U^\eps(t,t_0) \int_{t_0}^t \frac{d}{dt'}\left (
U^\eps(t_0,t') \UA^\eps(t',t_0) \right ) \, dt' \\
& = & -\frac{\ri}{\eps} U^\eps(t,t_0) \int_{t_0}^t
U^\eps(t_0,t') \left [ H(t') - \HA(t') \right ]\UA^\eps(t',t_0) \, dt' \\
& = & -U^\eps(t,t_0) \int_{t_0}^t
U^\eps(t_0,t') K(t') \UA^\eps(t',t_0) \, dt'.
\end{eqnarray*}
The idea is to rewrite $K(t)$ as a commutator, so that
$t \mapsto U^\eps(t_0,t) K(t) \UA^\eps(t,t_0)$ is the derivative of a function
(up to negligible terms), and an integration by parts gives the required estimates.
The proof proposed here is an extension of the proof presented in \cite[Chapter~2]{Teufel}
in the case when several pieces of the discrete spectrum are considered independently.
It would also have been possible to use
the twiddle operation introduced in~\cite{ASY87}, which is, in some sense, the inverse
operation of the commutator with the Hamiltonian.

\paragraph{Construction of the function used in the commutator.}
Consider $-\infty < t \leq 0$ such that $f(t) > 0$ (for compactly supported switching functions,
this means that $t$ is in the interior of the support). Define
\[
F(t) = -\frac12 \left ( \sum_{j=1}^{N+1} F_j(t) + G_j(t) \right ),
\]
with, for $1 \leq j \leq N$,
\begin{eqnarray}
  \label{eq:expression_Fj}
  F_j(t) & = & \frac{1}{2\ri\pi} \oint_{\Gamma_j(t)} P_j^\perp(t) R(z,t) \dot{R}(z,t) \, dz, \\
  \label{eq:expression_Gj}
  G_j(t) & = & \frac{1}{2\ri\pi} \oint_{\Gamma_j(t)} \dot{R}(z,t) R(z,t) P_j^\perp(t) \, dz,
\end{eqnarray}
where
\[
R(z,t) = (H(t)-z)^{-1},
\qquad
\dot{R}(z,t) = \frac{d}{dt} \left[ (H(t)-z)^{-1} \right] = -R(z,t) \frac{dH(t)}{dt} R(z,t),
\]
and $\Gamma_j(t)$ is a contour enclosing $E_j(t)$ and no other
element of the spectrum (such a contour exists in view of
Assumption~\ref{ass:gap}). For $j=N+1$, we denote by
$\Gamma_{N+1}(t)$ a contour enclosing all the first $N$ eigenvalues
$E_k(t)$, $k=1,\dots,N$, and separated from the remainder of the
spectrum (such a contour exists in view of
Assumption~\ref{ass:global_gap}), and define
\begin{eqnarray}
  \label{eq:expression_FN+1}
  F_{N+1}(t) & = & -\frac{1}{2\ri\pi} \oint_{\Gamma_{N+1}(t)} \left( \sum_{k=1}^N P_k(t) \right)^\perp
  R(z,t) \dot{R}(z,t) \, dz, \\
  G_{N+1}(t) & = & -\frac{1}{2\ri\pi} \oint_{\Gamma_{N+1}(t)} \dot{R}(z,t) R(z,t)
  \left( \sum_{k=1}^N P_k(t) \right)^\perp \, dz.
\end{eqnarray}
By definition of the contours,
\[
-\frac{1}{2\ri\pi} \oint_{\Gamma_j(t)} R(z,t) \, dz = P_j(t), \quad 1 \leq j \leq N,
\]
and
\[
-\frac{1}{2\ri\pi} \oint_{\Gamma_{N+1}(t)} R(z,t) \, dz = \sum_{k=1}^N P_k(t) = P_{N+1}^\perp(t).
\]
Besides, in view of the continuity of $t \mapsto E_j(t)$ for all $1 \leq j \leq N$,
it is possible to use contours which are locally constant in time, \ie for a given
$t > -\infty$ such that $f(t) > 0$,
there exists a (small) time interval $(t-\tau,t+\tau)$ and a contour
$\Gamma_j^t$ such that
\[
\forall s \in (t-\tau,t+\tau),
\quad
-\frac{1}{2\ri\pi} \oint_{\Gamma_j^t} R(z,s) \, dz = P_j(s)
\]
for $1 \leq j \leq N$, a similar result holding for $j = N+1$.
Using such locally constant contours, the time
derivative of the contour integral defining the projector can be restated as a contour integral
of the time derivative of the resolvent:
\[
-\frac{1}{2\ri\pi} \oint_{\Gamma_j(t)} \dot{R}(z,t) \, dz = \frac{dP_j(t)}{dt},
\quad 1 \leq j \leq N,
\]
and
\[
-\frac{1}{2\ri\pi} \oint_{\Gamma_{N+1}(t)} \dot{R}(z,t) \, dz
= \sum_{k=1}^N \frac{dP_k(t)}{dt} = -\frac{dP_{N+1}(t)}{dt}.
\]

\paragraph{Boundedness of $F$.}
The operator $F(t)$ is bounded. To see this, we first rewrite
$F_j$ ($1 \leq j \leq N$) as
\begin{equation}
  \label{eq:alternative_expression_Fj}
  F_j(t) = P_j^\perp(t) R(E_j(t),t) P_j^\perp(t) \frac{dP_j(t)}{dt}.
\end{equation}
Indeed, using the expression~\eqref{eq:expression_Fj} of $F_j$,
\begin{eqnarray*}
  & & F_j(t) - P_j^\perp(t) R(E_j(t),t) P_j^\perp(t) \frac{dP_j(t)}{dt} \\
  & & = \frac{1}{2\ri\pi} \oint_{\Gamma_j(t)} P_j^\perp(t) (R(z,t)-R(E_j(t),t)) P_j^\perp(t)
  \dot{R}(z,t) \, dz \\
  & & = -\frac{1}{2\ri\pi} \oint_{\Gamma_j(t)} P_j^\perp(t) (R(z,t)-R(E_j(t),t)) P_j^\perp(t) R(z,t)
  \dot{H}(t) R(z,t) \, dz. \\
\end{eqnarray*}
When the contour encircles closely enough $E_j(t)$,
\[
\| R(z,t) \| \leq \frac{2}{\delta_j(t)}.
\]
Using the resolvent identity, it follows
\begin{eqnarray*}
  \| P_j^\perp(t) (R(z,t)-R(E_j(t),t)) P_j^\perp(t) R(z,t) \|
  & = & |z-E_j(t)| \cdot \| R(z,t) P_j^\perp(t) R(E_j(t),t) P_j^\perp(t) R(z,t) \| \\
  & \leq & \frac{8|z-E_j(t)|}{\delta_j(t)^3}. \\
\end{eqnarray*}
Then, the difference
\[
\left \|
\oint_{\Gamma_j(t)} P_j^\perp(t) (R(z,t)-R(E_j(t),t)) R(z,t)
P_j^\perp(t) \dot{H}(t) R(z,t) \, dz
\right \| \leq C \frac{f'(t)}{\delta_j(t)^3} |\Gamma_j(t)|
\]
can be made arbitrarily small by decreasing the radius of the
contour $\Gamma_j(t)$, with a constant $C$ depending on the relative
$H_0$-bound of $V$.

From the expression \eqref{eq:alternative_expression_Fj}, and the
bound $\| P_j^\perp(t) R(E_j(t),t) P_j^\perp(t) \| \leq
\delta_j(t)^{-1}$, it holds finally
\[
\| F_j(t) \| \leq \frac{\| \dot{P}_j(t) \|}{\delta_j(t)} 
\leq C \frac{f'(t)}{f(t)},
\]
where we recall that both $f$ and $f'$ are non-negative.
This shows that $F_j(t)$ is a bounded operator when $f(t) > 0$.
A similar bound holds for $G_j$.

The terms $F_{N+1}(t), G_{N+1}(t)$ require a different treatment.
In this case, the uniformity of the gap between the $N$ eigenvalues
encircled by $\Gamma_{N+1}(t)$, and the remainder of the spectrum may be used
to construct a contour $\Gamma_{N+1}(t)$ such that
\[
\forall z \in \Gamma_{N+1}(t), \quad \| R(z,t) \| \leq \frac{4}{\Delta(t)}.
\]
This can be done by ensuring that the contour remains far away enough from the remainder
of the spectrum, while still being at a finite distance of the first $N$ eigenvalues.
In particular, it is possible to construct a contour intersecting
the real axis at a point $\gamma$ such that
$|\gamma-E_N(t)| \geq \Delta(t)/4$ and
\[
\inf \Big \{ |\gamma-E|, \ E \in \sigma(H(t)) \backslash \{ E_1(t),\dots,E_N(t) \} \Big \} \geq
\Delta(t)/4.
\]
Then,
\begin{equation}
  \label{eq:bound_FN+1}
  \| F_{N+1}(t) \| = \left \| \frac{f'(t)}{2\ri\pi} \oint_{\Gamma_{N+1}(t)}
  \left ( \sum_{k=1}^N P_k(t) \right)^\perp R(z,t)^2 V R(z,t) \, dz \right \|
  \leq C \frac{f'(t)}{\Delta(t)^3},
\end{equation}
and so $F_{N+1}$ is bounded since $\Delta(t) \geq \Delta^\ast > 0$ and $f'$ is bounded.
A similar bound holds for $G_{N+1}$.

In conclusion,
\begin{equation}
  \label{eq:bound_on_F}
  \|F(t)\| \leq C_F \,  \frac{f'(t)}{f(t)},
\end{equation}
for some constant $C_F$ independent of $t$.

\paragraph{Computation of the commutator.}
It is easily seen that $F(t)$ maps the Hilbert space $\mathcal{H}$ to $D(H_0)$.
The commutator $[H(t),F(t)]$ can then be defined as an unbounded operator with domain
$D(H(t)) = D(H_0)$. For a given $1 \leq j \leq N$, it holds, using
the commutation property $P_j^\perp(t) H(t) = H(t) P_j^\perp(t)$,
\begin{eqnarray*}
\left[H(t), F_j(t) \right] & = &
\frac{1}{2\ri\pi} \oint_{\Gamma_j(t)} [ H(t), P_j^\perp(t) R(z,t) \dot{R}(z,t)] \, dz \\
& = &  \frac{1}{2\ri\pi} \oint_{\Gamma_j(t)} [ H(t)-z, P_j^\perp(t) R(z,t) \dot{R}(z,t)] \, dz \\
& = & \frac{1}{2\ri\pi} \oint_{\Gamma_j(t)} P_j^\perp(t) \dot{R}(z,t)
- P_j^\perp(t) R(z,t) \dot{R}(z,t) (H(t)-z) \, dz\\
& = & -P_j^\perp(t) \frac{dP_j(t)}{dt} + P_j^\perp(t) \left ( \frac{1}{2\ri\pi}
\oint_{\Gamma_j(t)} R(z,t)^2 \, dz \right ) \dot{H}(t) \\
& = & -(\id - P_j(t)) \frac{dP_j(t)}{dt},
\end{eqnarray*}
following the proof of Theorem~2.2 in~\cite{Teufel}.
Similar computations show
\[
\left[H(t), G_j(t) \right] = \frac{dP_j(t)}{dt} (\id - P_j(t)).
\]
Finally, for $1 \leq j \leq N$,
\[
[H(t), F_j(t)+G_j(t)] = \left [ P_j(t),\frac{dP_j(t)}{dt} \right ].
\]
In the same way,
\[
[H(t), F_{N+1}(t)+G_{N+1}(t)] = -\left [ P_{N+1}(t),\frac{dP^\perp_{N+1}(t)}{dt} \right ]
= \left [ P_{N+1}(t),\frac{dP_{N+1}(t)}{dt} \right ].
\]
Since
\[
K(t) = -\sum_{j=1}^{N+1} P_j(t) \frac{dP_j(t)}{dt} =
-\frac12 \sum_{j=1}^{N+1} \left [ P_j(t),\frac{dP_j(t)}{dt} \right ],
\]
it holds
\begin{equation}\label{Eq:HFK}
[H(t),F(t)] = K(t).
\end{equation}

\paragraph{Integration by parts.}
Consider now $-\infty < t_0 < t \leq 0$ such that $f(t_0) > 0$ (hence $f(t) > 0$ since
$f$ is non-decreasing).
Define
\[
\mathcal{K}(t) = -\ri \eps \, U^\eps(t_0,t) F(t) U^\eps(t,t_0).
\]
Then
\[
\mathcal{K}'(t) = U^\eps(t_0,t) [H(t),F(t)] U^\eps(t,t_0) -
\ri \eps U^\eps(t_0,t) F'(t) U^\eps(t,t_0).
\]
In view of (\ref{Eq:HFK}), the difference between the
evolution operators is rewritten as
\begin{eqnarray}
  \label{eq:estimate_when_gap}
  U^\eps(t,t_0) - \UA^\eps(t,t_0) & = & -U^\eps(t,t_0) \int_{t_0}^t
  U^\eps(t_0,t') K(t') \UA^\eps(t',t_0) \, dt' \nonumber \\
  & = & -U^\eps(t,t_0) \int_{t_0}^t
  \left ( \frac{d\mathcal{K}(t')}{dt'} + \ri \eps U^\eps(t_0,t') \frac{dF(t')}{dt'}
  U^\eps(t',t_0) \right ) U^\eps(t_0,t') \UA^\eps(t',t_0) \, dt', \nonumber \\
\end{eqnarray}
so that, after an integration by parts for the term associated with $\mathcal{K}'$,
\begin{eqnarray}
  \label{eq:estimate_when_gap_2}
  \left \| U^\eps(t,t_0) - \UA^\eps(t,t_0) \right \| & = &
  \left \| \int_{t_0}^t U^\eps(t_0,t') K(t') \UA^\eps(t',t_0) \, dt'\right \|\\
  & \leq &
  \| \mathcal{K}(t)\| + \| \mathcal{K}(t_0)\|
  + \eps \int_{t_0}^t \left \| F' \right \|
  + \left \| \int_{t_0}^t \mathcal{K}(t') \frac{d}{dt'} \left (
  U^\eps(t_0,t') \UA^\eps(t',t_0)\right ) \, dt'\right \|
  \nonumber \\
  & \leq &
  \eps \left ( \| F(t)\| + \| F(t_0)\|
  + \int_{t_0}^t \| F'(t') \| \, dt'
  + \int_{t_0}^t \| F(t') \| \, \| K(t') \| \, dt' \right ).
\end{eqnarray}
The first two terms in the above equality are bounded with the bound
\eqref{eq:bound_on_F} on $F$. For the last one, we use again the
bound~\eqref{eq:bound_on_F} on $F$, and the fact that $K$ is
uniformly bounded (see~\eqref{eq:boundedness_of_K}), so that
\begin{equation}
  \label{eq:last-term_difference}
  \int_{t_0}^t \| F(t') \| \, \| K(t') \| \, dt' \leq C  \int_{t_0}^t \frac{(f')^2}{f}
  \leq \frac{C}{f(t_0)} \int_{t_0}^t (f')^2.
\end{equation}
We now turn to the central term. For $1 \leq j \leq N$, and using
\eqref{eq:alternative_expression_Fj},
\begin{eqnarray*}
  \int_{t_0}^t \| F_j'(t') \| dt'
  & \leq & \int_{t_0}^t \frac{\| \ddot{P}_j(t')\|}{\delta_j(t')} \, dt'
  + \int_{t_0}^t \| \dot{P}_j(t') \| \left \| \frac{d}{dt'} \left (
  P_j^\perp(t') R(E_j(t'),t') P_j^\perp(t')
  \right )\right \|\, dt' \\
  & \leq & \int_{t_0}^t \frac{\| \ddot{P}_j(t')\|}{\delta_j(t')} \, dt'
  + \int_{t_0}^t \frac{2\| \dot{P}_j(t') \|^2}{\delta_j(t')} \, dt' \\
  & & + \int_{t_0}^t \| \dot{P}_j(t') \| \, \| P_j^\perp(t') R(E_j(t'),t') V
  R(E_j(t'),t') P_j^\perp(t') \| \, f'(t') \, dt' \\
  & \leq & \int_{t_0}^t \frac{\| \ddot{P}_j(t')\|}{\delta_j(t')}
  + \frac{2\| \dot{P}_j(t') \|^2}{\delta_j(t')}
  + C f'(t) \, \frac{\| \dot{P}_j(t')\|}{\delta_j(t')^2} \, dt' \\
  & \leq & C' \int_{t_0}^t \left| \frac{f''(t')}{f(t')} \right|
  + 3 \, \frac{f'(t')^2}{f(t')} + \left( \frac{f'(t')}{f(t')} \right)^2 \, dt'\\
  & \leq & C ' \left ( \frac{1}{f(t_0)} \int_{t_0}^t \big( | f''| + 3 (f')^2 \big)
  + \frac{1}{f(t_0)^2} \int_{t_0}^t (f')^2 \right ), \\
\end{eqnarray*}
for some constants $C,C'> 0$ (related to the relative
$H_0$-bound of $V$).
Similar expressions can be obtained for $G_j$ ($1 \leq j \leq N$).
Straightforward estimates can be used for $F_{N+1},G_{N+1}$, following a treatment
similar to what was done to obtain~\eqref{eq:bound_FN+1}, upon deriving the terms
appearing in the contour integral:
\[
\| F'_{N+1}(t) \| \leq C \left ( \frac{|f''(t)|}{\Delta(t)^3}
+ \frac{f'(t)}{\Delta(t)^3} \sum_{k=1}^N \| \dot{P}_k(t) \|
+ \frac{f'(t)^2}{\Delta(t)^4} \right),
\]
with
\[
\| \dot{P}_k(t)\| = f'(t) \left\| \partial_\lambda \wP(f(t)) \right \|
\leq C \, f'(t).
\]
In conclusion,
\begin{equation}
  \label{eq:central-term_difference}
  \int_{t_0}^t \| F'(t') \| \, dt' \leq C \left ( \frac{1}{f(t_0)} \int_{t_0}^t
  \big( | f''| + (f')^2 \big)
  + \frac{1}{f(t_0)^2} \int_{t_0}^t (f')^2 \right ),
\end{equation}
for some constant $C > 0$.

\paragraph{Decomposition of the integral close to the degeneracy.}
In order to avoid the singularities when $f(t_0) \to 0$,
the difference of the unitary operators is separated into two contributions as
\beqn
U^\eps(0,t_0) - \UA^\eps(0,t_0) & = & -U^\eps(0,t_0) \int_{t_0}^T
U^\eps(t_0,t) K(t) \UA^\eps(t,t_0) \, dt \\
& & -U^\eps(0,t_0) \int_{T}^0
U^\eps(t_0,t) K(t) \UA^\eps(t,t_0) \, dt,
\eeqn
where $T$ is chosen such that $f(T) > 0$.
The first term is bounded using the straightforward estimate
\begin{equation}
  \label{eq:first_term}
  \left \| U^\eps(0,t_0) \int_{t_0}^T
  U^\eps(t_0,t) K(t) \UA^\eps(t,t_0) \, dt \right \|
  \leq C \int_{t_0}^T \sum_{k=1}^N \| \dot{P}_k(t) \| \, dt
  \leq C' \int_{t_0}^T f'(t) \, dt \leq C' f(T).
\end{equation}
For $t \in [T,0]$, $f(t) \geq f(T) > 0$ and there is a gap
proportional to $f(T)$ between the eigenvalues:
\[
\forall 1 \leq j \leq N, \quad \forall t \in [0,T],
\qquad \delta_j(t) \geq \alpha f(T),
\]
for some $\alpha > 0$.
The inequality~\eqref{eq:estimate_when_gap_2}, combined
with~\eqref{eq:bound_on_F}, \eqref{eq:last-term_difference}
and~\eqref{eq:central-term_difference},
allows to bound the second term as
\begin{equation}
  \label{eq:second_term}
  \begin{array}{l}
    \dps \left \| U^\eps(0,t_0) \int_{T}^0
    U^\eps(t_0,t) K(t) \UA^\eps(t,t_0) \, dt \right \|
    = \left \| \int_{T}^0
    U^\eps(T,t) K(t) \UA^\eps(t,T) \, dt\right \|
    \\[10pt]
    \dps \quad \leq C \eps \left( \frac{f'(0)}{f(0)} +
    \frac{f'(T)}{f(T)} +
    \frac{1}{f(T)} \int_T^0 \big( |f''| + (f')^2 \big) +
    \frac{1}{f(T)^2} \int_T^0 (f')^2 \right).
  \end{array}
\end{equation}
The limit $t_0 \to -\infty$ can then be taken in the above expressions. Moreover,
upon choosing $T$ small enough so that $f(T) = \eps^{1/3} \ll 1$, it follows,
adding \eqref{eq:first_term} and \eqref{eq:second_term}, and using the fact
that $f' \in \mathrm{L}^1((-\infty,0]) \cap \mathrm{L}^\infty((-\infty,0])$
and $f'' \in \mathrm{L}^1((-\infty,0])$,
\begin{equation}
  \label{eq:final_estimate_adiabatic}
  \left \| U^\eps(0,-\infty) - \UA^\eps(0,-\infty) \right \| \leq
  C \left ( f(T) + \eps \left (1 + \frac{1}{f(T)^{2}} \right ) \right )
  \leq 3 C \eps^{1/3}.
\end{equation}
This concludes the proof.


\subsection{Extensions}
\label{sec:extension}

The above proofs can be straightforwardly extended to the following
cases (see Section~\ref{sec:statement} for the notation).

\paragraph{Definition of the initial states when
  $\cP_0 V \cP_0$ has degenerate eigenvalues.}

Two changes should be made in the proofs presented in this paper:
(i) the estimate obtained in the adiabatic limit degrades;
(ii) more conditions are required to define the initial states.

Denote by $\cE_{0,i}$ the $M < N$ eigenspaces associated with the
eigenvalues of $\cP_0 V \cP_0$, set $n_i = \mathrm{dim}(\cE_{0,i})$,
and define
\[
\mathcal{N}_i = \Big \{ k \in \{ 1,\dots,N \} \, \Big | \, \varphi_{k,0} \in \cE_{0,i} \Big \},
\]
the set of indices corresponding to the $i$-th eigenspace of $\cP_0 V \cP_0$.
Of course,
\[
\sum_{i=1}^M n_i = N, \qquad \mathrm{Card}(\mathcal{N}_i) = n_i.
\]
In view of Assumption~\ref{ass:gap}, for any $(k,l) \in \mathcal{N}_i^2$,
$k \not = l$,
there exists an integer $p_{k,l} \geq 2$ and
an analytic function $e_{kl}(\lambda)$ such that
\[
E_k(\lambda) - E_l(\lambda) = \lambda^{p_{k,l}} \, e_{kl}(\lambda),
\qquad
e_{k,l}(0) \not = 0.
\]
Denote by $p_*$ the maximal integer for all couples $1 \leq k,l \leq N$. Then,
the final estimate \eqref{eq:final_estimate_adiabatic}
in the proof of the adiabatic limit reads
\[
\| U^\eps(0,-\infty) - \UA^\eps(0,-\infty) \| \leq
  C \left ( f(T) + \eps \left (1 + \frac{1}{f(T)^{2p_*}} \right ) \right )
  \leq 3 C \eps^{1/(2p_*+1)},
\]
which is indeed larger than the $\eps^{1/3}$ bound found in the case $p=1$
(no degeneracy of the perturbation restricted to $\cE_0$).

We now describe an iterative procedure which determines
the initial states in a unique manner, using the higher order equations
in the hierarchy \eqref{eq:eq_ordre_n}.
We start with the conditions of order 2. A necessary condition
for~\eqref{eq:eq_ordre_n} to have a solution is that its right-hand side belongs
to $\cE_0^\perp$. With \eqref{eq:1st_order_phi}, this requires, for all $1 \leq j, k \leq N$,
\begin{equation}
  \label{eq:determine_c1}
  \bra \varphi_{k,0}, \, V R_0 V \varphi_{j,0} \ket + E_{j,2} \delta_{j,k} + (E_{j,1}-E_{k,1})
  c_{j,k}^1 = 0,
\end{equation}
where $\delta_{a,b}$ is the Kronecker symbol.
In particular,
\[
\forall i \in \{ 1,\dots,M\}, \quad \forall (j,k) \in \mathcal{N}^2_i, \qquad \bra \varphi_{k,0},
V R_0 V \varphi_{j,0} \ket + E_{j,2} \, \delta_{j,k} = 0.
\]
Therefore,
$\{ \varphi_{j,0} \}_{j \in \mathcal{N}_i}$ has to be an eigenbasis of
$\cP_{0,i} V R_0 V \cP_{0,i}$
where $\cP_{0,i}$ denotes the projector onto $\cE_{0,i}$.
If $\cP_{0,i} V R_0 V \cP_{0,i}$ has non-degenerate eigenvalues,
the initial eigenfunctions $\{ \varphi_{k,0} \}_{k \in \mathcal{N}_i}$
are uniquely defined.

Otherwise, the procedure must be repeated.
Recall that there exists an integer $p_{*}$ such that after $p_{*}$ steps the degeneracy
has no further split (see the discussion at the beginning of this paragraph).
When the degeneracy is not permanent (see below for this case), 
this allows to determine the initial states in a unique manner.
See for instance~\cite{Hirschfelder69}.
In many practical cases however, degeneracy is never totally split
because V shares some symmetries with $H_0$. In this case, permanent degeneracy
has to be taken into account (see below).

\paragraph{Decomposition of the switching.}
In the case when \eqref{eq:condition_for_existence_ratio} is not
satisfied, \ie $\| P_j(0) - P(-\infty) \| = 1$ or
equivalently $\| P_j(0) \psi_j \| = 0$ (since the eigenspaces are assumed
to be one-dimensional), the switching should  be
done in several steps.
The intermediate steps can be chosen by
finding a finite number of values $\lambda_k \in
[0,1]$ ($k=1,\dots,N-1$), with $\lambda_0 = 0$ and $\lambda_N = 1$,
such that $\| \lP_j(\lambda_{k+1})-\lP_j(\lambda_k)\| < 1$. This is
possible since $\lP_j$ is continuous on the compact
interval $[0,1]$.

The initial state $\psi_0$ is evolved into a state
$\psi_1$ by switching from $H_0$ to $H_0 + \lambda_1 V$ as
\[
\psi_1 = \lim_{\eps_1 \to 0} \frac{U_{{\rm int},\lambda_1}^{\eps_1}(0,-\infty) \psi_0}
    {\left \langle \psi_0 \, \left | \,  U_{{\rm int},\lambda_1}^{\eps_1}(0,-\infty) \psi_0
      \right. \right \rangle},
\]
where the evolution operator
\[
U_{{\rm int},\lambda_1}^\eps(t,t_0) = \rme^{\ri t H_0/\eps} \, U_{\lambda_1}^\eps(t,t_0) \,
\rme^{-\ri t_0 H_0/\eps}
\]
is the following operator in the interaction picture:
\[
\ri \eps \frac{dU_{\lambda_1}^\eps(t,t_0)}{dt} = \Big ( H_0 + \lambda_1 f(t) V \Big )
U_{\lambda_1}^\eps(t,t_0), \quad U_{\lambda_1}^\eps(t_0,t_0) = \1.
\]
The state $\psi_1$ is then evolved into a state $\psi_2$ by switching $H_0 + \lambda_1 V$
to $H_0 + \lambda_2 V$ as
\[
\psi_2 = \lim_{\eps_2 \to 0} \frac{U_{{\rm int},\lambda_2,\lambda_1}^{\eps_2}(0,-\infty) \psi_1}
{\left \langle \psi_0 \, \left | \,  U_{{\rm int},\lambda_2,\lambda_1}^{\eps_2}(0,-\infty) \psi_0
      \right. \right \rangle},
\]
where the evolution operator
\[
U_{{\rm int},\lambda_2,\lambda_1}^\eps(t,t_0) = \rme^{\ri t H_0/\eps} \,
U_{\lambda_2,\lambda_1}^\eps(t,t_0) \, \rme^{-\ri t_0 H_0/\eps}
\]
is defined as the following operator in the interaction picture:
\[
\ri \eps \frac{dU_{\lambda_2,\lambda_1}^\eps(t,t_0)}{dt} = \Big (
H_0 + \lambda_1 V + (\lambda_2-\lambda_1) f(t) V
\Big ) U_{\lambda_2,\lambda_1}^\eps(t,t_0),
\quad
U_{\lambda_2,\lambda_1}^\eps(t_0,t_0) = \1.
\]
This construction is repeated until an eigenstate $\psi_N$ of $H_0 + V = H_0 + \lambda_N V$
is obtained. Notice that it is important to do the procedure sequentially.

\paragraph{Permanently degenerate eigenspaces.}
When there are permanently degenerate eigenspaces associated with one of the
eigenvalues $\lE_j(\lambda)$ or $E_j(t)$, the determination of the initial basis
can still be performed as it is presented in Section~\ref{sec:characterization}.
However, the argument leading to~\eqref{eq:1D_miracle}
in Section~\ref{sec:adiabatic_switching} cannot be extended as such
to the case when ${\rm Ran}\, \lP_j(0)$ is of dimension larger or equal to 2.
This is not a problem since $A(0,-\infty) \psi_j$ is still an
eigenvector of $P_j(0)$, and its phase can be removed upon considering
\[
\frac{U^\eps_{A, {\rm int}}(0,-\infty) \psi_j}{\bra \phi \rsep
  U^\eps_{A, {\rm int}}(0,-\infty) \psi_j \rangle}
= \frac{A(0,-\infty) \psi_j}{\bra \phi \rsep \left. A(0,-\infty) \psi_j \ket}
\]
for some fixed state $\phi$, provided the denominator is non zero.
In Theorem~\ref{thm:GLM}, the choice
$\phi = \psi_j$ is done, together with the assumption
$\bra \phi \rsep \left. A(0,-\infty) \psi_j \ket \not = 0$. This assumption could
in this specific case be translated into an assumption on $\| P_j(0) - P_j(-\infty) \|$,
but in general it should then be assumed that there exists $\phi \in \mathcal{H}$ such that
$\bra \phi \rsep \left. A(0,-\infty) \psi_j \ket \not = 0$.

\paragraph{Existence of finitely many crossings.}
The projectors being analytic, the Kato operator can still be defined when there are
eigenvalue crossings.
The main issue in extending the Gell-Mann and Low formula
to this case is therefore the proof of the adiabatic limit, which can however still be handled
with \cite[Corollary~2.5]{Teufel} since the crossings are regular (again, because
the eigenvalues are analytic).

\paragraph{Initial subspace composed of several degenerate spaces $\cE_0$, $\cE_1$,...}
In this case, the operator $V$ should be diagonalized in each subspaces, \ie
the self-adjoint finite-rank operators $\cP_j V \cP_j \big |_{\cE_j}$
are diagonalized in order to construct
a basis of $\cE_j$.
A global basis is then obtained by concatenation (direct sum).


\end{document}